# Coherent Growth and Characterization of van der Waals 1T-VSe$_2$ Layers on GaAs(111)B Using Molecular Beam Epitaxy

Tiancong Zhu[1,*#] Dante J. O'Hara[2,3,*Δ] Brenton A. Noesges[1], Menglin Zhu[4], Jacob J. Repicky[1], Mark R. Brenner[5,6], Leonard J. Brillson[1,5], Jinwoo Hwang[4], Jay A. Gupta[1], Roland K. Kawakami[1,2**]

[1]*Department of Physics, The Ohio State University, Columbus, OH 43210, USA*
[2]*Program in Materials Science and Engineering, University of California, Riverside, CA 92521, USA*
[3]*Materials Science Division, Lawrence Livermore National Laboratory, Livermore, CA 94550, USA*
[4]*Department of Materials Science and Engineering, The Ohio State University, Columbus, OH 43210, USA*
[5]*Department of Electrical and Computer Engineering, The Ohio State University, Columbus, OH 43210, USA*
[6]*Semiconductor and Epitaxy Analysis Laboratory, The Ohio State University, Columbus, OH 43210, USA*

**Abstract**

We report epitaxial growth of vanadium diselenide (VSe$_2$) thin films in the octahedrally-coordinated (1T) structure on GaAs(111)B substrates by molecular beam epitaxy. Film thickness from a single monolayer (ML) up to 30 ML is demonstrated. Structural and chemical studies using by x-ray diffraction, transmission electron microscopy, scanning tunneling microscopy and x-ray photoelectron spectroscopy indicate high quality thin films. Further studies show that monolayer VSe$_2$ films on GaAs are not air-stable and are susceptible to oxidation within a matter of hours, which indicates that a protective capping layer should be employed for device applications. This work demonstrates that VSe$_2$, a candidate van der Waals material for possible spintronic and electronic applications, can be integrated with III-V semiconductors via epitaxial growth for 2D/3D hybrid devices.

*Equal contributions

#Current affiliation: University of California, Berkeley

ΔCurrent affiliation: Naval Research Laboratory

**Corresponding Author

    Email:    kawakami.15@osu.edu
    Address:  191 W. Woodruff Ave.
               Department of Physics
               The Ohio State University
               Columbus, OH 43210
    Phone:    (614) 292-2515
    Fax:      (614) 292-7557

## 1. Introduction

The exploration of two-dimensional (2D) materials has received appreciable attention since the first isolation of monolayer graphene.[1–3] Being only a few atoms thick, the reduced dimensionality in 2D materials can strongly influence their electronic, optical and magnetic properties compared to their bulk analogues. For example, when reduced to monolayer thickness, the electrons in graphene behave as massless-Dirac fermions[2,3] while the electronic band structure in transition metal dichalcogenides (e.g. $MoS_2$, $WSe_2$) undergoes an indirect to direct gap transition with 100% valley selection with circular polarized light.[4–9] Other examples include the predicted tunable magnetism in hole-doped monolayer GaSe,[10] strong Ising pairing in superconducting $NbSe_2$ atomic layers,[11] and quantum spin Hall states in monolayer 1T'-$WTe_2$.[12,13] Controlling the material thickness and further studying how their properties depend on the number of layers is essential for understanding 2D materials.

Among the family of 2D materials, $VSe_2$ is of particular interest, because both its electronic and magnetic properties are significantly different for the bulk and in the monolayer limit.[14,15] Bulk $VSe_2$ may exhibit two polymorphs, which are the semiconducting 2H (trigonal prismatic) or metallic 1T (octahedral) phase. 1T-$VSe_2$ forms in a layered van der Waals structure, with each V atom octahedrally-coordinated by six surrounding Se atoms. Bulk 1T-$VSe_2$ undergoes a $(4 \times 4 \times 3)$ charge density wave (CDW) transition with $T_c = 110$ K.[16–20] However, the monolayer shows a $(\sqrt{7} \times \sqrt{3})$ CDW with altered-symmetry-type compared to the $(4 \times 4)$ CDW in the layers of the bulk.[21–24] On the other hand, although bulk 1T-$VSe_2$ is reported to be paramagnetic,[25,26] it has been shown to be ferromagnetic in its monolayer form with a Curie temperature above room temperature.[21,24] The drastic differences in bulk and monolayer 1T-$VSe_2$ make it an interesting material system to investigate how its properties change with increased thickness from the monolayer limit.

In this work, we demonstrate growth of large-area, high-quality 1T-$VSe_2$ on GaAs(111)B, a III-V direct gap semiconductor with favorable electronic, optoelectronic, and spintronic properties. Using molecular beam epitaxy (MBE), the film thickness is precisely controlled from single monolayers (ML) up to high thickness (30 ML). A combination of reflection high-energy electron diffraction (RHEED), x-ray diffractometry (XRD) and scanning transmission electron microscopy (STEM) confirms the formation of high quality, rotationally-oriented 1T-VSe$_2$ layers. X-ray photoelectron spectroscopy (XPS) measurements are conducted to confirm the chemical composition and to investigate the air stability of VSe$_2$ on GaAs. Furthermore, using scanning tunneling microscopy (STM), we confirm the atomic-scale structure of 1T-VSe$_2$ on GaAs(111)B and STM spectroscopy confirms the metallic nature of the monolayer film. These results establish a baseline for developing high quality thin films of van der Waals VSe$_2$ on III-V semiconductors, opening the door to novel electronic and spintronic devices based on 2D/3D hybrid structures.

## 2. Experimental Details

Vanadium diselenide growths are prepared by van der Waals epitaxy in a Veeco GEN930 chamber with a base pressure of $2\times10^{-10}$ Torr. Undoped GaAs(111)B substrates (AXT, single-side polished, 0.5 mm thick, $\pm$ 0.5° miscut, $1.4\times10^8$ $\Omega$-cm) are prepared by indium bonding to a 3" unpolished Si wafer then loading into the chamber and annealed at 400°C for 15 minutes under ultra-high vacuum (UHV) conditions ($1\times10^{-9}$ Torr) to remove water and gas adsorbates. The GaAs is then loaded into the growth chamber and exposed to a Ga flux (United Mineral & Chemical Corporation, 99.99999%, $1\times10^{-8}$ Torr) at 530°C for 2-4 minutes to remove the native oxides on the surface via Ga polishing.[27,28] The substrate is then annealed at 600°C under a Se flux ($1\times10^{-7}$ Torr) for 15 minutes to remove any excess oxide and terminate the surface with Se atoms then cooled to the growth temperature. Elemental Se (United Mineral & Chemical Corporation, 99.9999%) is evaporated from a standard Knudsen-type effusion cell with a typical cell temperature of 170°C and elemental V (ESPI Metals, 99.98%) is evaporated using a quad-rod electron-beam evaporator (MANTIS). The beam fluxes are measured using a nude ion gauge with a tungsten filament positioned at the sample growth position and the corresponding growth rate is calibrated based on nominal film thicknesses measured by *ex-situ* x-ray reflectometry (XRR, Bruker D8 Discover). *In situ* RHEED is used to monitor the growth and annealing procedures in real-time with an operation voltage of 15 kV. The VSe$_2$ growth is performed in an adsorption-limited growth regime at a fixed beam-equivalent pressure (BEP) Se:V ratio of ~50, where the excess Se re-evaporates. For *ex situ* characterization by XRD and STEM, samples are capped with amorphous Se or Te at room temperature to protect the surface from oxidation. For *in situ* characterization by XPS and STM, we mount the GaAs substrate onto a Omicron flag-style sample holder which is loaded into a Veeco uni-block adapter for MBE growth. After deposition, the sample on a flag-style holder is transferred to the separate XPS or STM chamber without air exposure via UHV suitcase. Due to the different sample mounting and its impact on growth temperature calibration, we rely on the RHEED patterns and their evolution during VSe$_2$ deposition to ensure that the growth occurs within the optimal temperature window.

XRD measurements are performed in a Bruker, D8 Discover system equipped with a Cu-K$\alpha$ 1.54 Å wavelength x-ray source. STEM is performed using a Thermofisher Titan STEM operated at 300 kV. Cross-sectional TEM specimens of VSe$_2$ were made using a focused Ga ion beam (Thermofisher Helios), which are subsequently cleaned using a low-energy Ar ion mill (Fishcione Nanomill) with a minimum beam energy of 500 eV. Atomic resolution high angle annular dark field (HAADF) images are acquired at the scattering range between 80 and 300 mrad, where the scattering intensity mostly depends on the atomic number of the elements.

XPS measurements are performed using a PHI VersaProbe 5000™ system equipped with a Scanning XPS Microprobe X-ray source (h$\upsilon_{Al\,K\alpha}$ = 1486.6 eV; FWHM ≤ 0.5 eV), and hemispherical energy analyzer with a pass energy of 23.5 eV and 0.05 eV step. To minimize the effects of charging, the XPS system is equipped with a two-stage sample surface neutralization system consisting of a 10 eV electron flood gun accompanied by a 10 eV Ar$^+$ ion beam. Photoelectrons are collected at a takeoff angle of 45°. Curve fitting is performed using asymmetric Gaussian for metallic and selenide-bonded vanadium core levels or Voigt line shapes oxidized vanadium.

STM measurements are performed with a CreaTec LT-STM/AFM system operating at 4.3 K in UHV with etched PtIr tips that are calibrated on clean Au(111) surfaces. We utilize conductive GaAs substrates (Si-doped GaAs(111)B from AXT, single-side polished, 0.5 mm thick, ± 0.5° miscut, (0.8~4)×10$^{18}$ cm$^{-3}$ carrier concentration) to allow conduction of the tip current. Following STM measurements, image analysis is performed with the WSxM software.[29]

## 3. Results and Discussion

We begin by determining the growth temperature window of VSe$_2$ on GaAs(111)B substrates for the highest quality thin films. GaAs(111)B is a substrate of choice for van der Waals epitaxy[30,31] due to the hexagonal symmetry of its surface layer and the widely-known passivation with Se atoms to form a GaSe-like surface. The deposition of VSe$_2$ on GaAs(111)B is illustrated in Figure 1 which shows RHEED patterns as a function of thickness and growth temperature. Optimal growth conditions are found at a substrate temperature of 170°C, Se:V flux ratio of ~50, and a growth of 1 ML in approximately five min. Figure 1a-d show the RHEED patterns for the passivated GaAs substrate and for 1 ML, 4 ML, and 6 ML VSe$_2$, respectively. The RHEED pattern of the VSe$_2$ remains streaky until approximately 30 MLs (~18.3 nm). Low temperature growths below 150°C result in a diffuse and dim RHEED pattern indicating an amorphous film from the large sticking coefficient and low atom mobility of Se. Figures 1e-f depict VSe$_2$ growth at temperatures higher than 200°C (the presented data is for 350°C), where the RHEED pattern is streaky initially but then quickly becomes spotty and modulates with azimuthal rotation after 1-2 monolayers (MLs) of deposition due to three-dimensional growth. Although growth temperature window is narrow, the RHEED patterns for 170°C growth (Figs. 1b-d) remain streaky and the intensity remains constant through larger thicknesses consistent with atomically smooth films. It is possible that increasing the Se flux can widen the temperature window to higher temperatures, but according to previous reports VSe$_2$ grown on highly-oriented pyrolytic graphite (HOPG) surfaces, a decrease in growth rate (higher chalcogen to metal flux) with a higher adatom mobility (increase in substrate temperature) can lead to large triangular islands without large-area homogenous coalescence across the whole substrate surface. To date, it is unclear if different island morphologies form when VSe$_2$ is deposited on GaAs. After deposition of 1 ML (Fig. 1b), the VSe$_2$ RHEED streaks coexist with the underlying GaAs(111). The inverse ratio of the RHEED spacing between the VSe$_2$ and GaAs(111) is measured to be 1.17, which is in agreement with the expected value of 1.19 for bulk in-plane lattice parameters of 1T-VSe$_2$(0001) ($a$ = 3.356 Å) and GaAs(111) ($a$ = 3.998 Å) despite the nearly 19% lattice mismatch. After ~2 MLs of growth, the RHEED streaks from the underlying GaAs disappear and the VSe$_2$ pattern remains streaky with six-fold rotational symmetry suggesting epitaxial alignment between the two materials, $[10\bar{1}0]_{VSe_2} \parallel [11\bar{2}]_{GaAs}$ and $[11\bar{2}0]_{VSe_2} \parallel [1\bar{1}0]_{GaAs}$. These results confirm that the VSe$_2$ grows fully relaxed on the GaAs surface, which is further confirmed by detailed *ex situ* characterization.

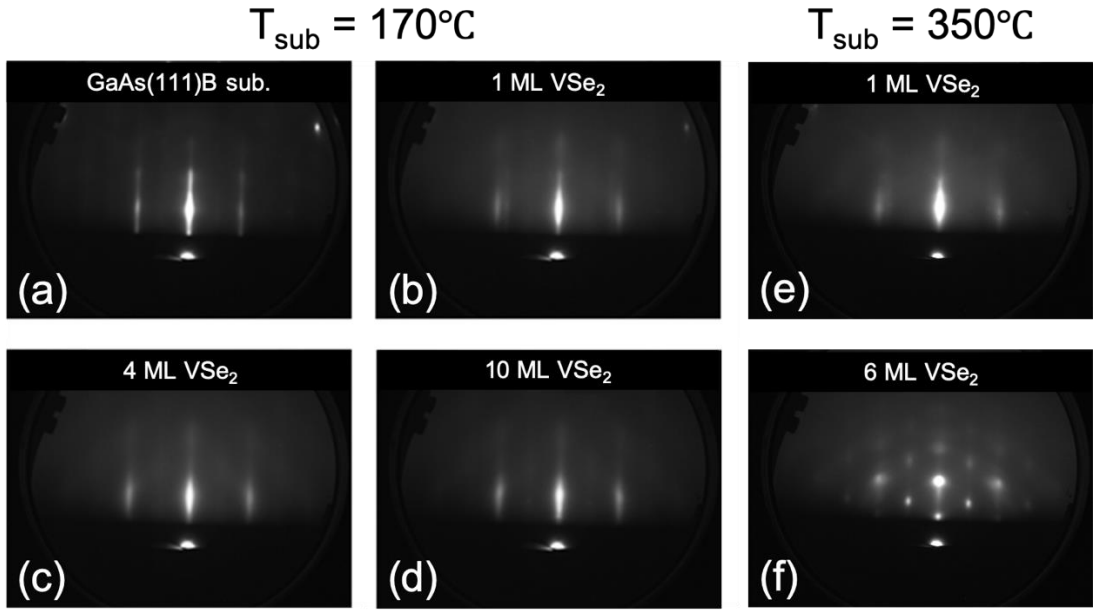

**Figure 1.** RHEED patterns of VSe$_2$ grown on GaAs(111) viewed along the $[11\bar{2}]$ of the substrate. (a-d) Streaky RHEED patterns of GaAs substrate, 1, 4, and 10 MLs of VSe$_2$ grown showing clear transition and streaky patterns confirming smooth and homogenous growth up to high thicknesses at a substrate temperature of 170°C. (e-f) Elevated substrate temperature growth of VSe$_2$ on GaAs showing three-dimensional growth by 6 MLs of deposition confirming a narrow temperature window.

We utilize x-ray diffraction (XRD) and cross-sectional STEM for detailed characterization of the interfacial and bulk structure, which is made possible by the ability to grow thick VSe$_2$ films. To determine the crystallographic orientation of the film, XRD measurements are performed on 30 ML VSe$_2$ films grown on GaAs (Fig. 2a). θ-2θ XRD scans show a weak (002) 1T-VSe$_2$ peak at 29.3° indicating a *c*-axis orientation of the film and a corresponding interplanar spacing of 6.1 Å in agreement with bulk crystals.[26,32,33] To further confirm the high crystalline quality of the film, atomic structure and interface quality, we perform cross-sectional high-resolution STEM (see Experimental Details). The high-angle annular dark field (HAADF) imaging in Fig. 2b shows clear van der Waals stacking from the abrupt interface between the grown VSe$_2$ film and the underlying GaAs substrate, confirms a high quality 1T structure with an epitaxial relationship of $[0002]_{VSe_2} \parallel [111]_{GaAs}$, and continuous growth of VSe$_2$ beyond 1 ML without any visible misfit dislocations propagating through the film or rotational disorder from different polytype phases (e.g. as observed in GaSe films[34]).

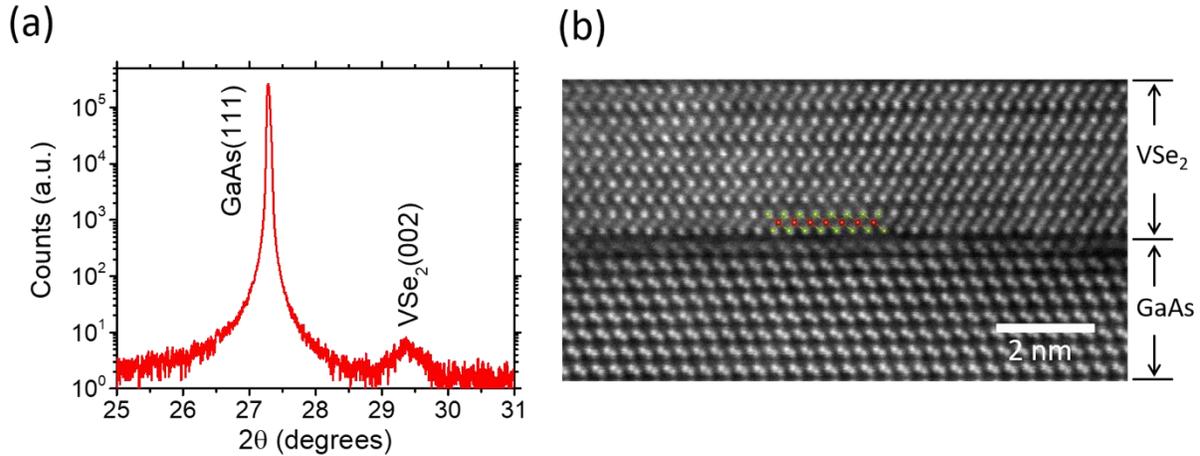

**Figure 2.** Structural characterization of thick VSe$_2$ films on GaAs(111) substrates with (a) showing XRD reflection of (00*l*) planes of 1T-VSe$_2$ orienting on GaAs(111) surfaces and (b) STEM image (viewed along <110> of GaAs) showing high quality VSe$_2$ layers grown on GaAs without the formation of misfit dislocations at the interface. Ball-and-stick model of 1T-VSe$_2$ is overlaid on the first atomic layer grown in the STEM image with red and green balls representing the V and Se atoms, respectively.

Air stability in ambient conditions is desirable for monolayer 2D materials. In order to study the air stability and chemical composition of our samples, we perform XPS measurements on monolayer VSe$_2$. Samples are first transferred from the MBE chamber to the XPS system by using a UHV suitcase. This allows us to measure the XPS spectrum of VSe$_2$ without any exposure to ambient conditions, which sets the baseline for the subsequent oxidation study. Figure 3 shows the XPS spectra around the V 2p region for as-grown monolayer VSe$_2$ and after 30 min or 15 h (total time) of air exposure. Distinct from studies of monolayer VSe$_2$ grown on HOPG where no surface oxidation was reported,[35] we observe the emergence of an oxygen (O) 1s peak. The primary component of the new O 1s feature is centered at 530.4 eV indicative of metal oxide formation.[36] In addition, the characteristic vanadium peaks show clear differences after air exposure (Figs. 3c,3d) with the V 2p$_{1/2}$ and V 2p$_{3/2}$ peaks shifting and splitting as a function of exposure time. After 30 min, both the V 2p$_{1/2}$ and V 2p$_{3/2}$ peaks lower in intensity while broadening and splitting into two peaks indicating the as-grown V-Se bonds are being replaced by a second bond type. The new identified peaks at 524.1 eV and 516.7 eV are indicative of a vanadium oxide structure forming at the surface. After 15 hrs. of air exposure, the V-Se characteristic peaks at lower binding energy vanish while the vanadium oxide peaks increase in intensity with centers at 523.5 eV and 516.3 eV. These two distinct peaks shifted to higher binding energy after prolonged air exposure compared to the as-grown peaks are consistent with the V-Se bonds being replaced by V-O bonds (similar shifts have been observed experimentally and analyzed theoretically for the oxidation of phosphorene[37]). These measurements provide evidence that monolayer $VSe_2$ is not oxidation-resistant when grown on GaAs surfaces, in contrast to previous reports of air-stability for MBE-grown monolayer $VSe_2$ on HOPG.[35] However, other factors such as grain size, morphology, and stoichiometry could also affect the result and further studies are required to understand the oxidation process. These results indicate that protective capping layers should be utilized for device applications.

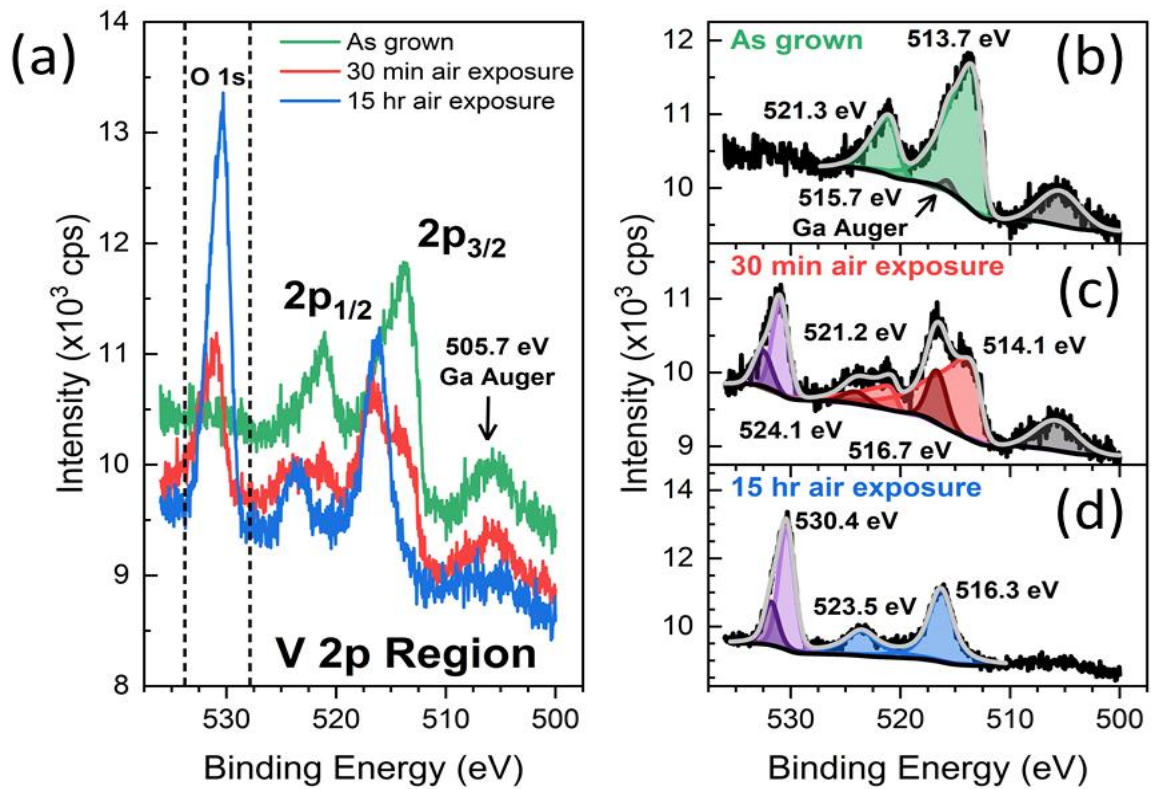

**Figure 3**. Air stability of monolayer VSe$_2$ on GaAs. (a) Overlaid XPS measurements of the V 2p characteristic region before (green) and after exposure to air (red and blue) showing changes in the characteristic peaks as a function of air exposure indicative of vanadium oxidation. (b-d) V 2p spectra separated as individual scans and deconvolved to evaluate the peak shift for V 2p$_{1/2}$ and V 2p$_{3/2}$ after air exposure using asymmetric and Voigt fittings for selenide and oxide states respectively. 30 min and 15 hr air exposures show an evolution to a fully oxidized surface and confirms the air instability of VSe$_2$ grown on GaAs.

The formation of 1T-$VSe_2$ is further confirmed by low-temperature STM. The surface of a 1 ML sample grown on GaAs is imaged through a topography measurement (Fig. 4a) revealing a well-ordered hexagonal atomic structure. From this, we can obtain an in-plane lattice constant of 3.44 ± 0.10 Å, close to previous reports of 3.356 Å for bulk $VSe_2$.[26,38,39] Figure 4b is a representative line profile showing the atomic corrugation. To investigate the local electronic structure of the $VSe_2$, we carry out STS measurements at low temperatures. The corresponding dI/dV spectra is shown in Fig. 4c, revealing the local density of states of the sample. The black curve represents the average dI/dV spectrum over several scans and the shaded red region represents the variation between different point spectra. The most notable feature of the dI/dV spectrum is the narrow dip at 0 mV with a minimum value near zero. A similar zero-bias anomaly has been observed in previous STM studies of monolayer $VSe_2$/HOPG at 150 K.[21] In that study, cooling from 150 K (no CDW) to 15 K induced CDW order which opened a gap of ~55 meV in the dI/dV spectrum. Our results on $VSe_2$/GaAs at 5 K are similar to $VSe_2$/HOPG at 150 K, where neither shows CDW order in the STM imaging at the given temperature (Fig. 4a) and the dI/dV spectra for both exhibit a narrow zero-bias anomaly (Fig. 4c).

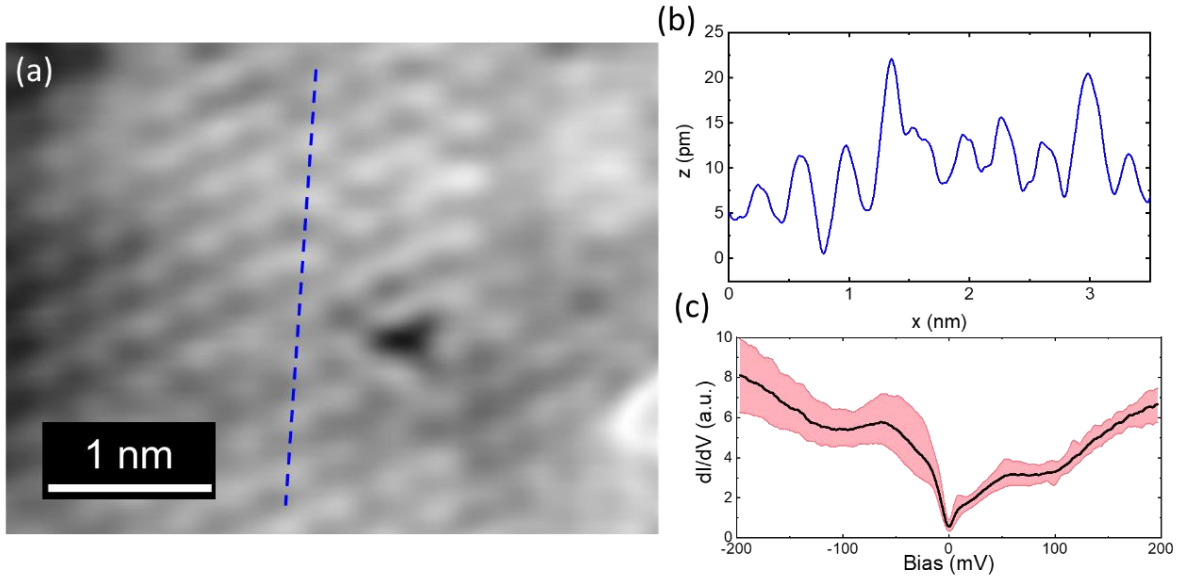

**Figure 4.** STM imaging and local electronic structure of 1 ML $VSe_2$ with (a) showing high resolution hexagonal lattice of $VSe_2$ and (b) Line profile of the atomic structure along the blue dashed line in (a). (c) dI/dV measurement confirming the local electronic structure is metallic.

## 4. Outlook

The realization of epitaxial growth of $VSe_2$ on a III-V semiconductor opens the door to novel 2D/3D hybrid structures for electronic, photonic, and spintronic devices. III-V semiconductors are employed in lasers, light-emitting diodes, high speed transistors, and optospintronic devices. Since room temperature ferromagnetism was reported for monolayer $VSe_2$,[21] its integration with III-V semiconductors is particularly interesting for spintronics. However, various studies report the presence or absence of ferromagnetism in $VSe_2$,[21,24,40–43] indicating that the origin of ferromagnetism is not well understood and suggesting a high sensitivity to sample preparation and microstructure. The integration of $VSe_2$ monolayers and thin films onto GaAs substrates enables a different approach to investigate magnetism, spin transfer, and spin dynamics in 2D/3D hybrid structures based on time-resolved Kerr rotation.[44,45] This technique utilizes a circularly-polarized pump pulse to generate electron spin polarization in GaAs and a time-delayed probe pulse to measure subsequent electron spin dynamics in GaAs and magnetization dynamics of the $VSe_2$. This provides an additional route to understand the origin of ferromagnetism in $VSe_2$ and investigate dynamical processes enabling hybrid 2D/3D spintronic devices.

## 5. Conclusions

In summary, we have demonstrated growth of atomically smooth and continuous thin films of 1T-$VSe_2$ on GaAs(111)B by MBE. Under optimal growth conditions, we can control the thickness from single ML up to 30 ML. Structural characterization by RHEED, XRD, STEM, and STM confirm the high quality of the interface, stacking, and surface structure. XPS experiments show poor air stability after measuring the films before and after exposure to ambient conditions. Electronic properties are investigated by low-temperature STM spectroscopy on monolayer films of VSe$_2$ on GaAs, which exhibits a metallic phase with a zero-bias anomaly.

**Acknowledgements**

We thank A. J. Bishop and C. H. Lee for support on the MBE growth. This work was primarily supported by the Department of Energy (DOE) Basic Energy Sciences under Grant No. DE-SC0016379. MZ, JH, BAN, and LJB acknowledge support from the Center for Emergent Materials, an NSF MRSEC, under Grant No. DMR-1420451. DJO recognizes support from Lawrence Livermore National Laboratory (LLNL) Lawrence Graduate Scholar Program and the GEM National Consortium Ph. D. Fellowship. Portions of this work were performed under the auspices of the U.S. Department of Energy by Lawrence Livermore National Laboratory under Contract DE-AC52-07NA27344. Electron microscopy was performed at the Center for Electron Microscopy and Analysis at The Ohio State University. DJO was supported by NRC/NRL while finalizing the manuscript.